\documentclass[sigconf,anonymous=false,screen]{acmart} 
\usepackage{booktabs}
\usepackage{array}
\usepackage{multirow}
\usepackage{fontawesome}
\usepackage{makecell}
\usepackage{float}

\newcommand{\insight}[1]{\noindent\faLightbulbO\hspace{3pt}\textit{#1}}

\AtBeginDocument{%
  \providecommand\BibTeX{{%
    \normalfont B\kern-0.5em{\scshape i\kern-0.25em b}\kern-0.8em\TeX}}}

\setcopyright{acmlicensed}
\copyrightyear{2025}
\acmYear{2025}
\acmDOI{XXXXXXX.XXXXXXX}

\acmConference[ ]{ }{ }{ }

\begin{document}

%\title{GenAI as a support tool for advanced undergraduate students}
\title{Student's Use of Generative AI as a Support Tool in an Advanced Web Development Course}


\author{Isaac Alpizar-Chacon}
\affiliation{%
 \institution{Utrecht University}
 \city{Utrecht}
 \country{The Netherlands}
}

\affiliation{%
 \institution{Instituto Tecnológico de Costa Rica}
 \city{Compus Cartago}
 \country{Costa Rica}
}
\email{i.alpizarchacon@uu.nl}

\author{Hieke Keuning}
\affiliation{%
 \institution{Utrecht University}
 \city{Utrecht}
 \country{The Netherlands}
 }
\email{h.w.keuning@uu.nl}


\begin{abstract}
Various studies have studied the impact of Generative AI on Computing Education. However, they have focused on the implications for novice programmers. In this experience report, we analyze the use of GenAI as a support tool for learning, creativity, and productivity in a web development course for undergraduate students with extensive programming experience. We collected diverse data (assignments, reflections, logs, and a survey) and found that students used GenAI on different tasks (code generation, idea generation, etc.) with a reported increase in learning and productivity. However, they are concerned about over-reliance and incorrect solutions and want more training in prompting strategies.
\end{abstract}





\maketitle

\section{Introduction}

Recent progress in artificial intelligence (AI) is changing Computing Education \cite{prather2023robots}. Generative AI (GenAI) can now, to a large extent, understand instructions in human language and create programming code. Many challenges and opportunities for CS education have been identified \cite{denny2024computing}, such as code generation, code explanations, academic misconduct, and over-reliance. Researchers have studied the performance and programming capabilities of such models in the context of CS1 and CS2 courses \cite{finnie2022robots,kazemitabaar2023studying}. Others have discussed the implications of GenAI for software development specifically \cite{ernst2022ai,daun2023chatgpt}. However, these implications have not been widely studied in advanced undergraduate courses. 

In this experience report, we describe how we incorporated GenAI into a advanced web development course taught to students in the penultimate year of a five-year information technology management program. We analyze the use of GenAI as a support tool for learning, creativity, and productivity in a context where students already have ample programming experience. 

We collected quantitative data (closed-ended survey questions), and qualitative data (assignments, reflections, logbooks, and open-ended survey questions). 
We aim to answer the following questions: 
1) How do advanced students use GenAI across different types of tasks/assignments?
2) What are their perceptions of GenAI use regarding learning, productivity, and ethics at different moments in the course?
and 3) What are the characteristics of their prompts?

We contribute to the existing body of literature on GenAI in computing education by describing the design of a GenAI-heavy course for upper-level students (Section \ref{sec:course}), providing insights into their behavior and views (Section \ref{sec:results}), and proposing suggestions for educators of such courses (Section \ref{sec:disc}).

\section{Related work} 

\paragraph{GenAI in computing education}

Many studies have provided a vision on the integration of GenAI into programming education \cite{10213396,denny2024computing}. 
Several earlier studies have focused on the teacher's perspective in general, studying their opinions and outlook on the future \cite{prather2023robots, lau2023ban}. 
A 2024 ITiCSE working group \cite{prather2024beyond} investigated what teachers were really doing in practice. They found that 75\% of the surveyed teachers believed that GenAI has an effect on programming competencies, and 30\% have changed their teaching as a result. They do this by means of focusing on different skills, installing policies on GenAI use, adapting lectures, and teaching students how to use it. 78\% of teachers do not explicitly disallow GenAI. The reasons that teachers incorporate GenAI include preparing students for the future, to support learning, and some teachers feel it is their responsibility. 

Students' general views on GenAI collected through surveys and interviews have also been studied extensively (e.g., \cite{prather2023robots, keuning2024students, smith2024early, choudhuri2024insights}).
\citet{keuning2024students} analyzed 264 survey responses of students from over 50 different computing courses, showing an increase in use over time. BSc students were more cautious of using GenAI, while MSc students are more comfortable to use it for programming, since that is not the learning goal of many advanced courses. 

\paragraph{GenAI in introductory programming}
The majority of studies have focused on the integration of GenAI into \textit{introductory} computing courses. \citet{10.1145/3626252.3630938} incorporated an AI-based suite of tools to assist students in learning the curriculum of Harvard University's CS50 course. Students could use the tools to get explanations of code, suggestions for improving code style, and a chatbot. Students highlighted the helpfulness, effectiveness, and reliability of the tools. \citet{10478015} explored students’ experiences with ChatGPT during a five-week introductory Java programming course. Students were allowed to use ChatGPT for all exercises.

They found that students viewed ChatGPT positively for learning programming concepts, though confidence in its use for implementation tasks was slightly lower. 
\citet{10.1145/3649217.3653584} described their CS1-LLM course, which focuses on skills such as explaining code, testing code, and decomposition—skills the authors argue are essential for producing software with GenAI assistance. Most students reported feeling that they could program effectively using GenAI, and a slight majority found the tools helpful for learning programming. However, some expressed concerns about becoming overly reliant on the tools or not fully grasping fundamental programming concepts. \citet{keuning2024goodbye} have proposed a novel CS1 course in which students experience a complete software development process, integrating GenAI in the different stages, but also emphasizing teamwork and connecting to real-life scenarios.

\paragraph{GenAI in other computing courses}

\citet{korpimies2024unrestricted} surveyed 41 students in a second year project course, in which students individually developed software. They could use GenAI freely, except for writing unit tests. 81\% used GenAI for tasks such as coding, documentation, bugfixing, explaining code, brainstorming, and learning course content. Students often mentioned that using GenAI saved them time, and made them more efficient, but also mentioned output problems. The authors found that students with `moderate' GenAI use spent more time than average on the project, while heavy users seemed to have slightly worse performance in the course. 

\citet{arora2024analyzing} explored how 411 undergraduates and graduates used GenAI during three programming projects in a Distributed Systems course. 
98\% of students used GenAI for a variety of tasks.  
Students also reported challenges similar to those by \citet{korpimies2024unrestricted}, such as obtaining relevant or accurate responses and difficulties distinguishing between correct and incorrect solutions.

\citet{choudhuri2024insights} interviewed (mostly senior) Software Engineering students about their GenAI use, and identified a broad spectrum of uses, as well as perceived benefits and challenges. They recommend careful integration of GenAI in courses and teaching students how to use it, for example through scaffolding.

\paragraph{Implications of students using GenAI}

Other studies focused on the implications of students' GenAI use, rather than their perceptions, giving more insight into the problems that may arise when students use GenAI. \citet{prather2024widening} identify a `widening gap' between the more skilled students who could use GenAI to even improve their work further, versus weaker students who struggled even more when using GenAI, also on a meta-cognitive level.
\citet{choudhuri2024how} investigate the effectiveness of ChatGPT in helping undergraduates (having limited knowledge of software engineering topics) with software engineering tasks (fixing code, removing code smells, and making pull requests), comparing students using ChatGPT to those that could only use `traditional' resources. They found no statistical difference in productivity (task correctness) or self-efficacy, but did find increased frustration levels.

\paragraph{This report} As in \citet{arora2024analyzing}, we also examine use, impact, and reflections on incorporating GenAI in an advanced course. However, unlike their study, which emphasizes students’ experiences with projects, our analysis covers a wider range of assignments incorporating GenAI and provides diverse insights across different stages of the course.
%: the beginning, middle, and end. 
Additionally, our course focuses on web development, a field where GenAI has demonstrated significant potential as valuable tools for enhancing productivity and streamlining coding tasks \cite{app142110048}. 

\section{Course Context} 
\label{sec:course}

\textit{Agile Web Application Development} (referred to hereafter as WEB) is a course that introduces foundational web development (HTML, CSS, JavaScript), web development frameworks (React, Angular, Node.js, Express, Django), and related topics (architecture, testing, security). The course is part of the Information Technology Administrator (ATI) program offered in Spanish at the Costa Rica Institute of Technology (TEC),
a university in Costa Rica specializing in STEM disciplines. ATI is a 5-year program divided into 10 semesters, leading to a licentiate degree: a Latin American professionally oriented, advanced undergraduate qualification, lower than a MSc degree but higher than a BSc degree. WEB is an elective typically taken in the 4th year, meaning that students already possess significant programming experience. The course consists of 16 weeks of lectures and 3 weeks of final assessments. Each week includes a 3-hour synchronous online instructor-led lecture, followed by 1 hour of offline lab work. Additionally, students dedicate self-study time for lab activities and other assignments. 

\subsection{Assessment}
The summative assessment includes four types of assignments, weighted as follows: weekly labs (45\%), homework (10\%), projects (40\%), and self-assessments (5\%). The 15 weekly labs introduce the students to additional technologies and tools essential for web development, such as GitHub Pages and ARIA. The four homework assignments focus on related web development topics, such as testing, through essays or presentations. In the three projects the students apply the learned frameworks to practical scenarios, such as building a ticketing system. Lastly, short weekly quizzes, both in-class and at-home, were conducted using Quizitor \cite{10.4018/IJMBL.318224}, a platform for assessments in classroom and online environments.

\subsection{GenAI integration}
% \usepackage{booktabs}
\begin{table}[tb!]
\caption{Course assignments with a GenAI component.}
\label{tab:assignments}
\small
\begin{tabular}{@{}llll@{}}
\toprule
\textbf{Assignment} &
  \textbf{Type} &
  \textbf{\begin{tabular}[c]{@{}l@{}}Timing\end{tabular}} \\ \midrule %  in \\ Course
\textsc{Lab 1} &
  Solo &
  \begin{tabular}[c]{@{}l@{}}Start (wk. 1)\end{tabular} \\ 
\textsc{Homework 1} &
  Solo  &
  \begin{tabular}[c]{@{}l@{}}Start (wk. 3)\end{tabular} \\ 
\textsc{Project 1} &
  Duo &
  \begin{tabular}[c]{@{}l@{}}Midway (wk. 7 \& 8)\end{tabular} \\ 
\textsc{Project 2} &
  Duo &
  \begin{tabular}[c]{@{}l@{}}Midway  (wk. 10 to 12)\end{tabular} \\ 
\textsc{Homework 3} &
  Solo  &
  \begin{tabular}[c]{@{}l@{}}Midway (wk. 12 \& 13)\end{tabular} \\ 
\textsc{Project 3} &
  Duo &
  \begin{tabular}[c]{@{}l@{}}Midway (wk. 13 to 17)\end{tabular} \\ 
\textsc{Logbook} &
  Solo &
  \begin{tabular}[c]{@{}l@{}}Throughout (wk. 1 to 17)\end{tabular} \\ 
\textsc{Survey} &
  Solo &
  \begin{tabular}[c]{@{}l@{}}End (wk. 16)\end{tabular} \\ \bottomrule
\end{tabular}
\end{table}

Following the introduction of GPT-3.5 and GitHub's Copilot in 2022, many challenges and opportunities of such tools in Computing Education have been identified \cite{becker2023programming,denny2024computing}. Given that WEB was a new course, first offered during the second semester of 2023 (July to November), it provided a unique opportunity to integrate GenAI tools into its learning activities.  The aim was not only to leverage GenAI as a support tool for mastering the course content but also to encourage students to reflect on and learn about using such tools in educational contexts. Students were allowed to use GenAI tools freely for all assignments, with specific tasks designed to promote critical reflection and deeper understanding of the role these tools play in both the learning process and the development workflow. The course was designed and taught by an experienced instructor with ten years of teaching experience.

In the first lecture, the instructor introduced resources and studies \cite{lau2023ban,kazemitabaar2023studying,Porter_Zingaro_2023,Shani_2023} on the use, impact, and limitations of GenAI for programming and software development to motivate the course's GenAI usage policy. No specific instructions on how to use GenAI were provided. However, in the second lecture, the instructor demonstrated the use of Copilot with a simple JavaScript example.

Several assignments were designed to either incorporate the use of GenAI or encourage reflection on its role and impact, see Table \ref{tab:assignments}. \textsc{Lab 1} required students to reflect on their expectations of using GenAI for learning, both in general and in the context of the course, through seven targeted questions (e.g., "What advantages do you see in using Artificial Intelligence tools in the learning process?" and "Are there situations where you prefer not to use them?"). \textsc{Homework 1} involved reading a paper on the opportunities, challenges, and implications of ChatGPT across multiple domains \cite{DWIVEDI2023102642}, followed by writing a summary and briefly discussing the importance of six out of the 43 described essays in the paper.
Group projects tasked students to build increasingly complex applications: a text adventure game using HTML5, CSS3, and JavaScript (\textsc{Project 1}), a React-based front-end meeting management system (\textsc{Project 2}), and a full-stack ticketing system with Angular and Express (\textsc{Project 3}). In these projects, students were required to describe whether and how they used GenAI tools, reflect on the process, and evaluate the benefits of using these tools for the project. In \textsc{Homework 3} students had to select a topic, 
use GenAI to explore it, and write a short essay incorporating web and scientific resources while documenting the GenAI-assisted discovery process. Also, students had to maintain a \textsc{Logbook}, recording their interactions with GenAI tools during self-study and assignments. Students also completed an end-of-course \textsc{Survey}, primarily based on a survey by a 2023 ITiCSE working group \cite{prather2023robots}. We translated the survey into Spanish, made minor adjustments, and added some questions focusing on learning and productivity.

\subsection{Participants}
All 14 students of the 2023 iteration of WEB passed the course and joined the study. Each student signed a consent form. They agreed that all GenAI-related assignments could be used for analysis and that the results, including anonymized quotes, may be used and shared in publications. The form also specified that participation was voluntary, participants had the right to withdraw at any time, responses would be kept confidential, stored on a secure institutional server, and the data would be deleted 10 years after the study.

Table \ref{tab:background} presents background information about the participants, taken from the \textsc{Survey}. To maintain anonymity, gender was not explicitly asked, but the participant list included 12 names traditionally associated with masculine identities and 2 with feminine identities. Additionally, based on the program year, we can deduce that all participants began their studies in 2020 or earlier. This means they completed their introductory programming courses well before the widespread use of GenAI tools.

\begin{table}[tb!]
\caption{Participants' year in the program (out of five), and prior web-programming experience and programming proficiency (self-estimated from 1-beginner to 5-advanced).}
\label{tab:background}
\setlength{\tabcolsep}{3.9pt} % Adjust horizontal spacing
\small
\begin{tabular}{@{}lllllllllllllll@{}}
\toprule
\textbf{Student} & 
  \multicolumn{1}{c}{1} &
  \multicolumn{1}{c}{2} &
  \multicolumn{1}{c}{3} &
  \multicolumn{1}{c}{4} &
  \multicolumn{1}{c}{5} &
  \multicolumn{1}{c}{6} &
  \multicolumn{1}{c}{7} &
  \multicolumn{1}{c}{8} &
  \multicolumn{1}{c}{9} &
  \multicolumn{1}{c}{10} &
  \multicolumn{1}{c}{11} &
  \multicolumn{1}{c}{12} &
  \multicolumn{1}{c}{13} &
  \multicolumn{1}{c}{14} \\ \midrule
\begin{tabular}[c]{@{}l@{}}\textbf{Program Year}\end{tabular}                          & 4 & 4 & 4 & 4 & 5 & 4 & 4 & 4 & 4 & 4 & 4 & 4 & 4 & 5 \\
\begin{tabular}[c]{@{}l@{}}\textbf{Web Experience}\\ (avg=2.5)\end{tabular}         & 1 & 4 & 3 & 3 & 3 & 3 & 2 & 3 & 3 & 1 & 2 & 3 & 2 & 2 \\
\begin{tabular}[c]{@{}l@{}}\textbf{Prog.} \textbf{Prof.}\\ (avg=3.6)\end{tabular} & 3 & 5 & 3 & 4 & 4 & 5 & 4 & 4 & 3 & 3 & 3 & 3 & 4 & 3 \\ \bottomrule
\end{tabular}
\end{table}

\section{Data analysis}

We conducted different analyses of the data collected throughout the course, translated into English (see Table \ref{tab:assignments}). One author, the course instructor, provided in-depth knowledge, while the other, uninvolved in the course, brought objectivity to the analysis.

\paragraph{First assignments} We follow a descriptive approach to report students’ reflections in \textsc{Lab 1} and their comments on the essays read in \textsc{Homework 1}, highlighting and emphasizing their most significant responses.

\paragraph{Projects}
For the three \textsc{projects}, we use the taxonomy from \citet{tufano2024unveiling}, who analyzed commits, pull requests, and issues, mined from software repositories. The taxonomy contains the types of development tasks that can be automated using ChatGPT. We use an extended version that adds some additional use cases from computing students \cite{keuning2024students}.

From the students' reflections on their GenAI use, we extracted the names of the tools used, deductively coded how they were used with categories from the taxonomy mentioned above, and inductively identified themes related to benefits. We conducted the analysis of the first project with two authors in a collaborative session. The remaining projects were coded by one author, and cases of doubt were discussed with another author.

\paragraph{Prompts}
To analyze the prompts and accompanying comments in the \textsc{Logbooks} and \textsc{Homework 3}, we categorize them in several dimensions, following \citet{alves2024givecodelog}'s approach. They analyzed the prompts used by first-year undergraduate students to solve programming problems. We adopted some of their categories and labels while extending others inductively based on the students’ interactions and comments. One author conducted the coding. 

\paragraph{Survey}
For the \textsc{Survey}, due to its length, we focus on a selection of questions. For open-ended questions, we adopt a descriptive approach, summarizing and highlighting the main themes.

\section{Experiences}
\label{sec:results}

In this section, we describe our findings and summarize the key insights at the end of each part, marked with light bulb icons \insight{}.

\subsection{GenAI views at the start of the course}\label{results:start}

To the \textsc{survey} question on how familiar the students were with AI programming tools before this course, 8 answered they used them regularly, 4 tried them a few times, and 2 stated they had heard about them but had not tried them. These responses highlight the diverse levels of familiarity at the start of the course. \textsc{Lab 1} encouraged students, following the first class, to reflect on how GenAI tools might impact their learning throughout the course. The majority expressed a positive attitude toward using GenAI for programming tasks, such as debugging. 
%Several students highlighted the speed and efficiency of finding relevant information. 
However, some maintained a more conservative perspective, emphasizing the importance of responsible use to avoid ``requesting complete solutions'' or ``relying entirely on these tools.'' Students highlighted several advantages of using GenAI tools in their learning process, including constant availability, faster response times, textual capabilities, and programming support. However, they also critically noted limitations, such as incorrect or misleading outputs, the risk of dependency, and reduced learning. Students also expressed privacy concerns, like sharing sensitive or personal data inadvertently and uncertainty about data handling and third-party access. When asked about rules or policies regarding using GenAI, some students assigned responsibility to teachers (e.g., creating GenAI-proof assignments) or individuals, emphasizing the need for self-monitoring to ensure they are still learning while using these tools. Others compared its regulation to traditional plagiarism policies or expressed more liberal views advocating for fewer restrictions. 

For \textsc{Homework 1}, students selected six short essays from a paper and discussed why the topics of these essays were important to them. Their choices spanned a wide range of topics, including essays related to education and critical thinking. Some students focused on essays aligned with their studies, such as ChatGPT’s role in the IT industry. Others chose domain-specific essays, exploring applications of ChatGPT in fields like banking services or tourism. Students also described why those topics were important. Many students emphasized the importance of AI literacy, including technical knowledge, highlighting ``the need to deeply understand how the technologies we use work.'' Others explored various ethical aspects, such as plagiarism and biases, or highlighted the importance of GenAI for learning and teaching, mentioning critical thinking, learning support, and curriculum changes, for example. 
Some students emphasized the significance of GenAI in the workplace, highlighting its role in driving digital transformation and enhancing productivity. Others focused on human-AI collaboration.

From the start of the course, we have the following insights:

\insight{Students acknowledged, at the beginning of the course, a range of potential benefits and risks of using GenAI tools for learning.}

\insight{Accountability and privacy are concerns, with some emphasizing personal responsibility and others assigning responsibility to teachers.}

\insight{Students recognized the diverse impact of GenAI across education, ethics, and the workplace.} 

\subsection{Using GenAI for web development projects}

\begin{figure}[t]
    \centering
    \includegraphics[scale=0.5]{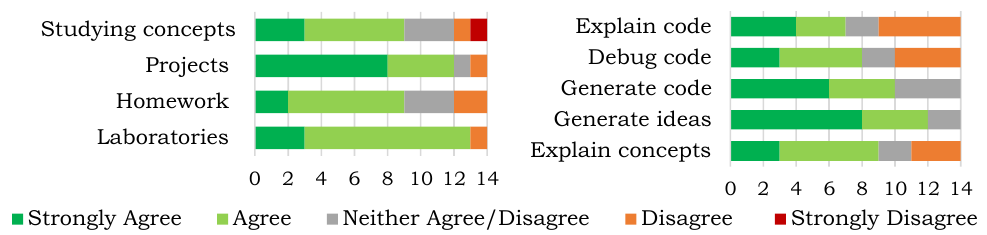}
    \caption{Answers to question `For this course, I regularly used GenAI when working on... (left) and to... (right)'.}
    \label{fig:use}
\end{figure}

Figure \ref{fig:use} shows how students reported using GenAI for various course components, and for what purposes.
The heaviest use is in the projects, and they used it most to generate code and ideas. We now discuss the 19 reflections on GenAI use for these projects.

Students used different GenAI tools, in a broad range of ways. As expected, they mostly use ChatGPT and Copilot, but also mention other generic AI tools (Bard, You.com) and more specific tools (Canva, Picsart).
All teams mention ChatGPT for all projects (except one reflection that did not mention tools). 
One duo mentioned that they used Bard for ``additional insight and alternative solutions to those offered by ChatGPT'', because it is ``connected to the internet.''

Many duos used GenAI for implementing features, either in a generic way by using Copilot to give code suggestions, or by asking ChatGPT for a specific solution. Some applications are for boilerplate code, such as getters/setters, CSS, or HTML tables.
Spotting and fixing bugs are also mentioned often. One duo even mentioned that it helped them to \textit{avoid} common mistakes: ``it alerted us to possible problems before they could become serious errors.'' %p3 11-14
As a new addition to the tasks of the taxonomy \cite{tufano2024unveiling}, we noticed that many students mentioned the generation of resources. \textsc{Project 1} particularly, building a text adventure, provided ample opportunity to be creative. Canva, Picsart, and Bing image generator were used to generate images, and ChatGPT to enhance the storyline.
Another new use case we observed was the use of GenAI in tasks related to requirements engineering.
One duo mentioned using GenAI to ``provide guidance to define the functionalities of our platform [...] defining its purpose [...], giving us suggestions on how to present the functionalities in a clearer and more attractive way for potential users.'' % p2 5/10
A duo also used ChatGPT to provide examples of similar platforms, to serve as inspiration for their own solution. %p3 11-14
These uses match the open-ended nature of the project. Although the core functionalities were given, students had freedom to choose the specific application area. For example, the ticketing system could be for concerts, but also for a conference.

Documentation was not mentioned often. One duo generated code comments using the Mintlify Doc Writer VS Code extension.
A duo also mentioned using GenAI for finding the right API: ``ChatGPT helped us understand that the API we were using was not the best fit [...]. Although he did not provide us with a specific alternative, his analysis of the bugs and limitations of the original API gave us a clearer understanding of our needs, allowing us to investigate and find a more suitable solution on our own.'' %p2 5/10

Students often mention that they asked GenAI tools \textit{how} to approach some task, implying that it is more of a learning experience than just generating the solution. However, this is a subtle difference that we observe, and should be confirmed by more closely looking at the actual prompts that they write.

Most duos are positive and mention benefits such as time-saving and therefore increased productivity, higher quality of their final product, and a more enjoyable experience with less frustration. A duo specifically mentioned ``the time saved allowed us to focus on UX refinement.'' 
While some reflections solely raved about the advantages of using GenAI, several also discussed downsides. 
For example, a duo tried using GenAI for exporting and validating JSON, however, the results were insufficient or had to be much adjusted. 
Another duo said that ``for some components it is a greater waste of time and a source of frustration to be talking to the tool [...] I want [...] to simply use my own knowledge and research time to create exactly what is needed.'' 
Other students were also careful: ``we noticed that in certain cases it provided the exact answer [...] but other times it presented errors that nevertheless pointed to the eventual solution [...]. We still see this tool as something useful that should be used carefully and tested extensively." They used GenAI as a ``last resort'' in cases where the given documentation did not suffice.
One duo concluded that ``AI is useful for starting projects or understanding topics quickly", but also commented it should be used sparingly for code generation ``as it is not always functional or contextualized with the rest of the application." Correcting it can take as much time as creating it from scratch. 

We conclude with our main insights:

\insight{Students can use GenAI for support with tasks aimed at creativity and user experience, which they are usually less comfortable with, and not part of their learning goals.}

\insight{Students use GenAI extensively when the course encourages this.}

\insight{Students use GenAI for a broad range of tasks, including code generation, requirements engineering, testing, documentation, and UI. }

\subsection{Prompting skills} 

Table \ref{tab:prompts} shows the analysis of 144 prompts (97 from \textsc{Logbooks} and 47 from \textsc{Homework 3}). Since students did not consistently use the \textsc{Logbooks}, the data represents only a sample of their interactions. 

Students used various GenAI tools, with ChatGPT being the most used. Most interactions were done in Spanish, the students' native language. However, some prompts were in English. One student noted that ``I chose to write it in English because most of the information is available in this language.'' Students used prompts for a variety of tasks: exploring a concept/topic (e.g., discussing advantages and disadvantages), requesting code, retrieving information (e.g., list of CSS frameworks), and generating or enhancing text (e.g., a story or a description). Most prompts were straightforward requests (e.g., ``What is GitHub Pages"), though approximately 31\% employed additional strategies. For instance, one student incorporated a definition of microservices in web development before asking ``How to implement, test, and maintain microservices?'' Some prompts specified constraints, such as the programming language or defining the desired output (e.g., ``Provide three examples''). Only one prompt included example code. One student observed that ``prompts can be optimized to retrieve all the information at once,'' while another crafted a prompt solely to ``generate context within the AI'' for subsequent interactions. Additionally, students rated 63\% of responses as highly useful, reporting some problems. For code generation, some noted the need to modify the generated code and consult additional resources. Interestingly, 2 students used the same prompt in different tools, probably to compare results. 

In \textsc{Homework 3}, students observed that certain tools (e.g., ChatGPT) do not reference the information they generate: ``I had no way to verify its truthfulness.'' However, the same student leveraged the tool’s response to identify keywords for searching academic articles on the selected topic. Another student used Perplexity
because it ``provides the sites from which the information was taken and used to generate its responses.'' Finally, multiple students recognized that using GenAI provided a basic knowledge foundation, enabling them to continue their research using other resources. 

From this analysis, we see two main insights:

\insight{Only some students display advanced prompting skills}.

\insight{Students exhibit critical thinking by evaluating the responses}.

\begin{table}[]
\small
\caption{Prompt analysis results (N=144).}
\label{tab:prompts}
\begin{tabular}{@{}ll@{}}
\toprule
\textbf{Feature} & \textbf{Values}                                                                                                                                \\ \midrule
Tool             & \begin{tabular}[t]{@{}l@{}}ChatGPT (82\%), Copilot (6\%), Perplexity (4\%), Bard (2\%), \\ You.com (2\%), Bing (1\%), Other (3\%)\end{tabular} \\
Language         & Spanish (86\%), English (14\%)                                                                                                                 \\
Type &
  \begin{tabular}[t]{@{}l@{}}Explore concept (24\%), Ask for code (16\%), Explain concept \\ (16\%), Explain how (15\%), Retrieve Information (10\%), \\ Help with error (4\%), Generate text (4\%), Edit code (3\%), \\ Create Image (3\%), Enhance Text (2\%), Other (3\%)\end{tabular} \\
Strategy         & \begin{tabular}[c]{@{}l@{}}Context (13\%), Restrictions (17\%), Examples (0.7\%)\end{tabular}                    \\
Usefulness       & High (63\%), Medium (27\%), Low (4\%), Unknown (6\%)                                                                                           \\
Problem          & \begin{tabular}[t]{@{}l@{}}Incorrect/incomplete answer (6\%), Code not working (4\%), \\ General answer (2\%), Other (4\%)\end{tabular}        \\ \bottomrule
\end{tabular}
\end{table}

\subsection{Reflections at the end of the course}

\begin{figure}[tb!]
    %\centering
    \includegraphics[scale=0.52]{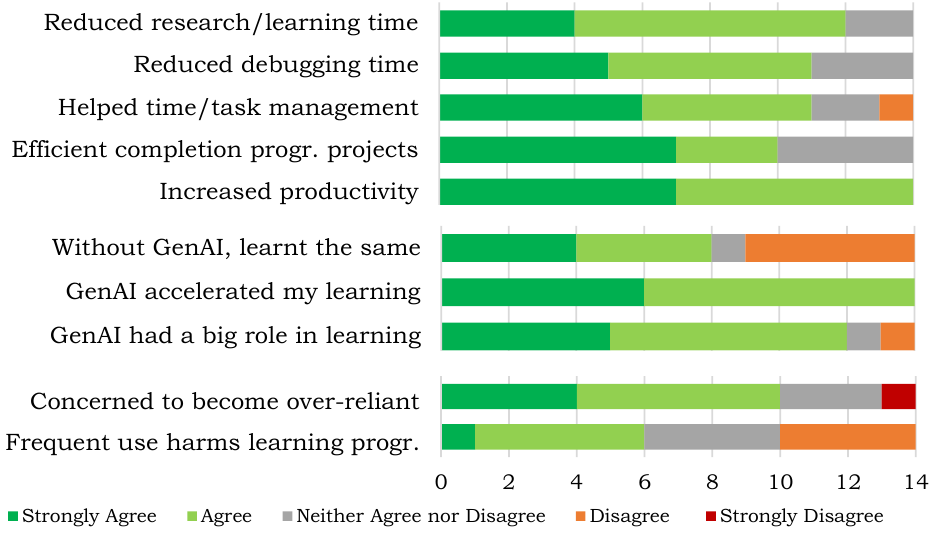}   
    \caption{Answers to questions about the perceived effect of using GenAI on productivity (top), learning (middle), and concerns (bottom).}
    \label{fig:effects_productivity}
\end{figure}

Figure \ref{fig:effects_productivity} shows the positive impact of GenAI tools on productivity (top) and learning (middle). Students indicated improvement in writing documents, finding information faster in a centralized platform, and, mostly, ease and speed for writing code. One student noted ``They have helped me reduce the time spent creating the base code and solving problems, allowing me to focus on developing the functions or aspects that require my creativity and thought.'' In terms of learning, many highlighted how GenAI tools helped them understand the topics and concepts, mentioning ``they allowed me to learn about topics that, on my own, would have been extremely difficult for me.'' For programming-related tasks, students noted benefits such as ``Understanding in detail the steps for different programming procedures'' and ``Seeing practical examples in code.'' However, two students expressed challenges, mentioning that such tools occasionally interrupted their learning process. One noted, ``At times, it has made the process more difficult since the tool provides a result that does not work correctly.''

At the beginning of the course (Section \ref{results:start}), some students expressed concerns about becoming overly reliant on GenAI. As shown in the bottom of Figure \ref{fig:effects_productivity}, this apprehension was shared by the majority of students by the end of the course. Additionally, slightly less than half of the students believed that frequent use of GenAI could negatively impact their ability to learn programming. However, despite these concerns, when asked about the continuation of using GenAI in the course, all students supported its use. They mentioned its importance with statements such as  ``it is pointless to go against it'', and ``in the workplace, they are absolutely essential nowadays.'' Students also gave suggestions to improve the GenAI's integration in the course: greater student participation in class to assess topic comprehension, teaching prompting, and discussing the answers from GenAI tools.

Finally, regarding ethics, all but one student felt confident identifying situations where GenAI should not be used due to its limitations. Moreover, the majority agreed that the diverse course assignments made them more aware of the ethical implications of using GenAI. Students also highlighted various ethical aspects, including dependency on the tools, copying and pasting solutions or code without understanding them, decisions around using fully GenAI-generated code, and privacy concerns. Notably, one student reflected ``Sometimes I didn’t read what it was telling me and just copied and pasted it, so I reflected that I shouldn’t rely so much on AI. Instead, I should verify what it tells me and use it as support.''

From the final reflections, we see the following insights:

\insight{Students expressed concerns about the risk of becoming overly reliant on GenAI tools after using them during the course.}

\insight{Students reported that using GenAI tools positively impacted their productivity and learning throughout the course.}

\insight{Students supported using GenAI in the course but emphasized the need for additional activities, such as training on effective prompting.}

\section{Discussion} 
\label{sec:disc}

\paragraph{Insights}
At the start of the course, only some students expressed concerns about over-reliance on GenAI tools, but the number of students with these concerns increased significantly by the end. They also highlighted in the \textsc{survey} the need for teachers to have more control in assessing topic comprehension. This suggests that, despite being advanced students, they struggle with self-regulating their learning and require more support when using GenAI tools. Students interacted with GenAI throughout the course, but fewer than one-third of the analyzed prompts employed advanced strategies that could have led to better responses. Additionally, only a few students used GenAI tools that provide references, while others admitted they could not verify the accuracy of the answers. These experiences, along with students explicitly requesting training in prompting strategies at the course’s end, highlight the need for better guidance on selecting appropriate GenAI tools for different situations and using them effectively.

\paragraph{Assignments}
\textsc{Lab 1} and \textsc{Homework 1} encouraged the students to reflect early on using GenAI in an educational context. \textsc{Homework 3} demonstrated an effective use of GenAI as a supporting tool for gathering initial knowledge, which could then be verified by consulting additional sources. The \textsc{Projects} allowed students to leverage GenAI for diverse tasks, highlighting both its capabilities and limitations. On the other hand, the \textsc{logbooks} were used inconsistently, with students documenting only some of their interactions with GenAI. In the \textsc{survey} students requested changes to the \textsc{logbooks}, citing the the cumbersome nature of recording all interactions. In the second iteration of WEB, held in the first semester of 2024, the same GenAI policy was implemented with minor adjustments. First, the \textsc{logbook} was replaced by a reflection at the end of each assignment where GenAI was used, focusing on its use and impact on learning. Additionally, for \textsc{Homework 1}, students were required to select at least one essay about ethics from the available ones in the paper \cite{DWIVEDI2023102642}. Data was also collected during this iteration; however, with only four students in the course, drawing general conclusions was infeasible. Despite this, two themes emerged: students expressed satisfaction with the open policy on GenAI use and reiterated concerns about becoming overly reliant.

\paragraph{Limitations}
This study primarily examined students’ perceptions, as reflected in surveys and written parts of assignments, which presents a limitation. Other studies have shown that perceived benefits, such as saving time, or learning, do not always match with reality \cite{prather2024widening}. Similarly, since WEB did not include exams, it was not possible to directly assess students’ learning of course topics in an environment without the use of GenAI. Students could also have been dishonest on their use of GenAI, however, because the course encouraged the use, this would be limited.

\section{Conclusion} 

This experience report provided an in-depth analysis on the use of GenAI in different types of assignments and tasks by advanced undergraduate students with prior programming experience. We gathered both quantitative and qualitative data during a 19-week web development course, in which the unrestricted use of GenAI was permitted. Various assignments were incorporated to encourage students to reflect on their use of GenAI.

Students used GenAI regularly during the course across assignments (labs, homework, and projects) and tasks (explaining concepts, generating code, generating ideas, etc.). Students reported that productivity increased due to GenAI usage and that it played a significant role in their learning. However, students are concerned about over-reliance and errors in the generated code. Additionally, students would like to learn how to use GenAI better.

As future work, we aim to address students’ concerns (e.g., fear of over-reliance), incorporate additional activities into the course (e.g., training on effective prompting), and continue monitoring the impact of GenAI tool usage on advanced students.

\clearpage
\bibliographystyle{ACM-Reference-Format}
\bibliography{references}

\end{document}